\documentclass{appolb}
\usepackage{graphicx,color}
\usepackage{amsmath}
\usepackage{amssymb}

\begin{document}
\global\long\def\ave#1{\left\langle #1 \right\rangle }
\global\long\def\absol#1{\left| #1 \right|}
\global\long\def\mev{{\rm \, MeV}}
\global\long\def\gev{{\rm \, GeV}}
\global\long\def\dpp#1{\frac{d^{3}#1}{(2\pi)^{3}}}
\global\long\def\mh{\hat{\mu}}

\title{Fluctuations and the QCD Phase Diagram}%

\author{Volker Koch
\address{Nuclear Science Division\\
Lawrence Berkeley National Laboratory\\
Berkeley, CA, 94720, USA }
\\ 
{Adam Bzdak
}
\address{AGH University of Science and Technology\\
Faculty of Physics and Applied Computer Science\\ 
30-059 Krak\'ow, Poland}
}
\maketitle
\begin{abstract}
In this contribution we will discuss how the study of various fluctuation
observables may be used to explore the phase diagram of the strong
interaction. We will briefly summarize the present study of experimental
and theoretical research in this area. We will then discuss various
corrections and issues which need to be understood and applied for
a meaningful comparison of experimental measurements with theoretical
predictions. This contribution is dedicated to Andrzej Bialas on the
occasion of his $80^{\mathrm{th}}$ birthday.
\end{abstract}

\section{Introduction }
\label{sec:intro}

Soon after the discovery of QCD \cite{Fritzsch:1973pi}, and following
the realization that QCD exhibits asymptotic freedom \cite{Gross:1973id,Politzer:1973fx},
is was recognized that QCD predicts a new high temperature phase of
weakly interacting quarks and gluons, termed the Quark Gluon Plasma
\cite{Collins:1974ky,Cabibbo:1975ig,Shuryak:1977ut}. The existence
of a new phase was confirmed in the first calculations using the lattice
formulation of QCD, initially for pure $SU(2)$ gauge theory \cite{Creutz:1980zw,Creutz:1979dw}.
Over the years, as Lattice QCD methods have been refined to allow
for continuum extrapolated calculations with dynamical quarks at the
physical masses, it has been found that the transition from hadrons
to partons at vanishing net baryon density is an analytic cross over
\cite{Aoki:2006we}. At the same time many model calculations suggested
that at vanishing temperature but large baryon density there might
be a first order transition \cite{Stephanov:2004wx}. This first order
phase transition will end at a critical point, the location of which
is not really constrained by any model calculations let alone Lattice
QCD, which, due to the fermion sign problem can only explore regions
of small net-baryon chemical potential, $\mu_{B}/T\lesssim1$.

The potential presence of a first order phase co-existence region
together with a critical point has motivated a dedicated experimental
program at RHIC, the so-called RHIC Beam Energy Scan (BES). Experimentally,
regions of higher baryon density can be reached by lowering the beam
energy where some of the projectile and target baryons are stopped
at mid-rapidity. The study of fluctuations play an important role
in the quest to experimentally explore the QCD phase diagram. Both,
the second order phase transition associated with a critical point
and the first order transition give rise to characteristic fluctuation
pattern. Of course the system produced in a heavy ion collision is
of finite size and evolves in time which smoothens the singular structures
associated with phase transitions. However, fluctuation measurements
are still helpful in this case, because, as we shall discuss below,
fluctuations are related to derivatives of the free energy. For example
cumulants of the baryon number are given by derivatives with respect to
the baryon chemical potential, $\mu_{B}$, etc. Therefore, the measurement
of cumulants of a sufficient high order will allow to explore experimentally
if there are any ``wiggles'' in the free energy, which may be associated
with some phase changes.

In addition to thermal fluctuations there are many other sources and
types of fluctuations. On the most fundamental level there are quantum
fluctuations, which arise if we measure several non-commuting observables.
In heavy-ion collisions, we encounter fluctuations and correlations
related to the initial state of the system, fluctuations reflecting
the subsequent evolution of the system, and trivial fluctuations
induced by the experimental measurement process. Initial state fluctuations
are driven by, e.g., inhomogeneities in the initial energy and baryon number deposition.
These fluctuations are quite substantial, and are reflected, for example,
in higher harmonics of the radial flow field. 

In this contribution, we will concentrate on thermal fluctuations,
which, away from some phase transitions, are typically small, suppressed
by $1/\sqrt{N}$ where $N$ is the average number of particles in
the volume considered. We will also be concerned with fluctuations
originating in the measurement. These need to be understood, controlled
and subtracted in order to access the dynamical fluctuations which
tell as about the properties of the system. 

In experiment fluctuations are most effectively studied by measuring
so-called event-by-event (E-by-E) fluctuations, where a given observable
is measured on an event-by-event basis and its fluctuations are studied
for the ensemble of events. Alternatively, one may analyze the appropriate
multi-particle correlations measured over the same region in phase
space \cite{Bialas:1999tv}.

This contribution is organized as follows. We will first provide a
short review on thermal fluctuations and how they can be addressed,
e.g., by Lattice QCD. We will then discuss various corrections which
need to be applied to the data and (model) calculations. We will close
with a discussion of the recent preliminary measurement of net-proton
cumulants by the STAR collaboration. Finally, we wish to dedicate this
contribution to Andrzej Bialas on the occasion of his
$80^{\mathrm{th}}$ birthday.

\section{Fluctuations of a thermal system}
\label{sec:2}

The system created in a ultra-relativistic heavy ion collision reaches,
to a very good approximation, thermal equilibrium
(see e.g. \cite{Braun-Munzinger:2015hba} for a recent review). 
Thermal fluctuations are typically characterized
by the appropriate cumulants of the partition function or, equivalently,
by equal time correlation functions which in turn correspond to the
space-like (static) responses of the system. In the following we will
concentrate on cumulants of conserved charges, such as baryon number
and electric charge. To this end, we will work within the grand-canonical
ensemble, where the system is in contact with an energy and ``charge''
reservoir. Consequently, the energy and the various charges are only
conserved on the average with their mean values being controlled by
the temperature and the various chemical potentials. As far as heavy
ion reactions are concerned, the grand canonical ensemble appears
to be a good choice as long as one considers a sufficiently small
subsystem of the entire final state. In addition, as discussed e.g.
in \cite{Andronic:2014zha}, the final state hadron yields are very
well described by a grand canonical thermal system of hadrons.

Fluctuations of conserved charges are characterized by the cumulants
of susceptibilities of that charge. Given the partition function of
the system with conserved charges $Q_{i}$ 
\begin{align}
Z={\rm Tr}\left[\exp\left(-\frac{H-\sum_{i}\mu_{i}Q_{i}}{T}\right)\right]
\end{align}
the susceptibilities are defined as the derivatives with respect to
the appropriate chemical potentials. In case of three flavor QCD the
conserved charges are the baryon number, strangeness and electric
charge, $\left(B,S,Q\right)$, and we have 
\begin{align}
\chi_{n_{B},n_{S},n_{Q}}^{B,S,Q}\equiv\frac{1}{VT^{3}}\frac{\partial^{n_{B}}}{\partial\mh_{B}^{n_{B}}}\frac{\partial^{n_{S}}}{\partial\mh_{S}^{n_{s}}}\frac{\partial^{n_{Q}}}{\partial\mh_{Q}^{n_{Q}}}\ln Z,\label{eq:fluct:susz_genreal}
\end{align}
where $\mh_{i}=\mu_{i}/T$ is the reduced chemical potential for charge
$i$. Since the pressure is given by $P=(T/V)\ln(Z)$, the above susceptibilities
also control its Taylor expansion for small values of the various
chemical potentials. For example 
\begin{align}
\frac{P\left(T,\mu_{B}\right)}{T^{4}}=\frac{P\left(T,\mu_{B}=0\right)}{T^{4}}+\sum_{n}c_{n}\,\left(\mu_{B}/T\right)^{n},\label{fluct:eq:pressure_mu}
\end{align}
with the expansion coefficients given by $c_{n}=\frac{\chi_{n}^{B}}{n!}$.
Such a Taylor expansion is employed in order to determine the QCD
equation of state for small chemical potentials \cite{Allton:2002zi,Gavai:2008zr,Borsanyi:2012cr}
from lattice QCD, since the Fermion sign problem does not allow for
a direct calculation. Let us next discuss two examples which illustrate
how the study of fluctuations and correlations provide insight into
the structure of QCD matter.

\subsection*{Net Charge Fluctuations }
\label{sec:fluct:charge}

One example are the fluctuations of the net electric charge. In Refs.
\cite{Jeon:2000wg,Asakawa:2000wh} it has been realized that the electric
charge $q$ of particles contributes in square to the fluctuations
of the net-charge, 
\begin{eqnarray}
\ave{\left(\delta Q\right)^{2}}=q^{2}\ave{\left(\delta N\right)^{2}}=q^{2}\ave N,
\end{eqnarray}
where in the last step we assumed the particles to be uncorrelated.
Thus, cumulants of the net-charge are sensitive to the fractional
charge of quarks in a quark gluon plasma. To remove the dependence
on the system size it is convenient to scale the charge variance by
another extensive quantity, such as the entropy, 
\begin{align}
R & =\frac{\ave{\left(\delta Q\right)^{2}}}{S}\label{R-def}
\end{align}
A simple estimate using Boltzmann statistics gives \cite{Jeon:2000wg,Jeon:2003gk}
\begin{align}
R_{QGP}=\frac{1}{24}
\end{align}
for a two flavor quark-gluon plasma whereas for a gas of massless
pions we obtain 
\begin{align}
R_{\pi}=\frac{1}{6}.
\end{align}
In other words, due to the fractional charges of the quarks and the
increased entropy due to the presence of gluons, the charge
fluctuations per entropy in a QGP is roughly a factor four smaller
than that in a pion gas at the same temperature. In reality the hadronic
phase contains more particles than pions, and, taking into account
hadronic resonances, the charge variance per entropy is reduced by
about 30\% which leaves about a factor three difference between a
hadronic system and a QGP. Also, a system of constituent quarks,
without any thermal gluons leads to a ratio of charge fluctuation to
entropy similar to a hadron resonance gas \cite{Bialas:2002hm}.
Finally, it is worth pointing out that similar arguments
have been utilized to identify the fractional charges in a quantum
Hall system as well as the double charge of cooper pairs in measurements
of shot noise \cite{shot_quantum_hall,shot_cooper}.

\begin{figure}[t]
\begin{center}
\includegraphics[width=0.55\textwidth]{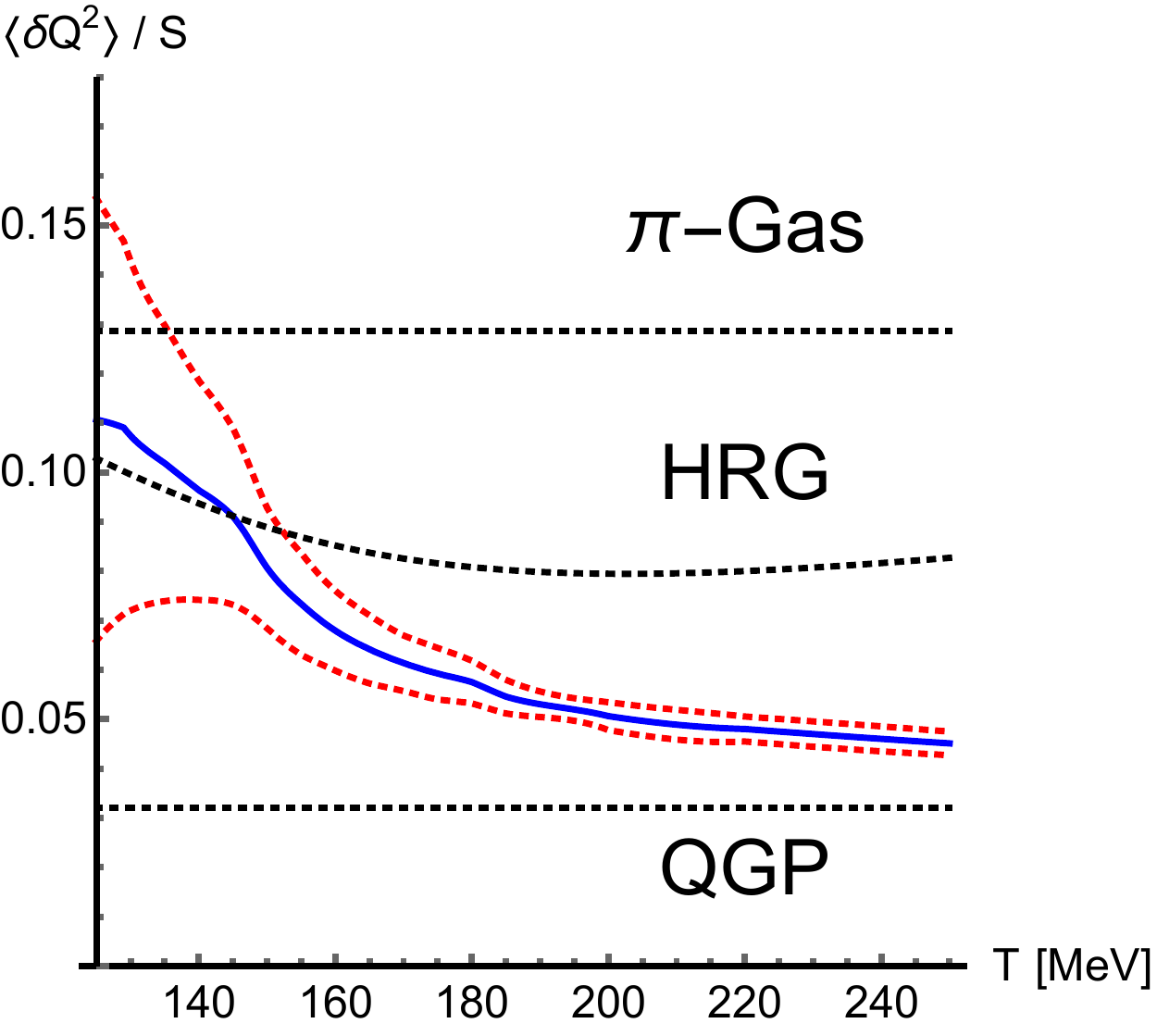} 
\end{center}

\caption{Net-charge variance per entropy, $R$, as a function of temperature
from 2+1 flavor lattice QCD with physical quark masses. The red-dashed
line indicate the uncertainty. Results for $\protect\ave{\left(\delta Q\right)^{2}}$
are from \cite{Borsanyi:2011sw} and the entropy is extracted from
\cite{Borsanyi:2010cj}. The dashed horizontal lines indicate the
results for a massless pion gas, a hadron gas as well as a non-interacting
QGP with three flavors of massless quarks. }
\label{fig:fluct:charge_fluct_lattice} 
\end{figure}

Since the above ratio, Eq. (\ref{R-def}), contains only well defined
thermal observables, it can be determined using lattice QCD methods,
thus accounting for all possible correlations, the presence of strange
quarks etc. In Fig.~\ref{fig:fluct:charge_fluct_lattice} we show
lattice QCD results for the net-charge variance per entropy based
on the calculations for the net-charge variance from \cite{Borsanyi:2011sw}
and for the entropy density from \cite{Borsanyi:2010cj}. We also
show the results for a free pion gas and a QGP with three flavors
of massless quarks, both using the proper quantum statistics, as
well as that for a hadron resonance gas. We see that the hadron resonance
gas agrees well with the lattice results for temperatures up to, $T\lesssim160\mev$,
which is close to the pseudo-critical or cross-over temperature of $T_{pc}=154\pm8\mev$.
For temperatures in the range of $160\mev\lesssim T\lesssim250\mev$
the lattice calculations are in between the resonance gas prediction
and that of a non-interacting QGP, indicating that some of the correlations
leading to resonance formation are still present in the system. With
few exceptions, this trend is seen for most quantities which have
been calculated on the lattice, such as energy density, cumulant ratios,
etc. Good agreement with the hadron resonance gas up to the cross-over
temperature, followed by a rather smooth transition to a free QGP
which takes place over a temperature interval of approximately $\Delta T\sim100\mev$,
where the correlations slowly disappear. As we will show in the next
example, some of these correlations, namely those between the various
quark-flavors, can be explored explicitly by studying so called mixed
flavor or ``off-diagonal'' cumulants.

\subsection*{Correlations between quark flavors }

\begin{figure}[t]
\begin{center}
\includegraphics[width=0.7\textwidth]{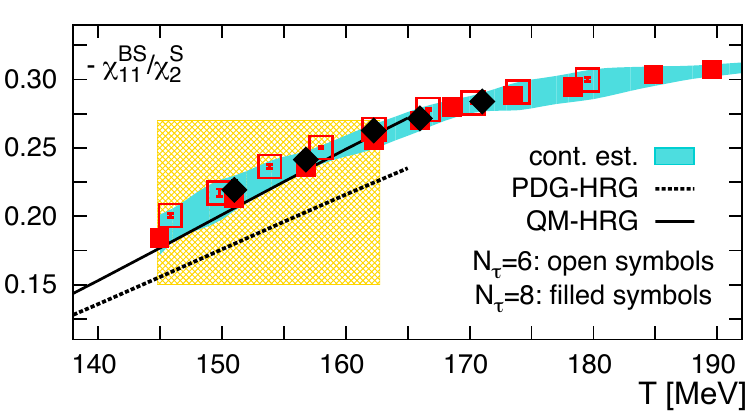} 
\end{center}

\caption{Lattice QCD results for $-\frac{\chi^{B,S}_{11}}{\chi^{S}_{2}}=\frac{1}{3}C_{BS}$
together with results from hadron resonance gas with and without extra
strange mesons. Figure adapted from \cite{Bazavov:2014xya}.}
\label{fig:fluct:cbs} 
\end{figure}

Let us start by considering the co-variance between strangeness and
baryon number, $\ave{\delta B\delta S}\sim\chi_{1,1}^{B,S}$. To illustrate
the sensitivity of this co-variance to correlations among quarks,
let us again compare a non-interacting QGP with a non-interacting
hadron resonance gas (HRG). In the QGP strangeness is carried exclusively
by baryons, namely the strange quarks, whereas in a HRG strangeness
can also reside in strange mesons. Therefore, baryon number and strangeness
are more strongly correlated in a QGP than in a hadron gas, at least
at low baryon number chemical potential, where mesons dominate. To
quantify this observation, Ref.~\cite{Koch:2005vg} proposed the
following quantity 
\begin{align}
C_{BS}\equiv-3\frac{\ave{\delta B\delta S}}{\ave{\delta S^{2}}}=1+\frac{\ave{\delta u\,\delta s}+\ave{\delta d\,\delta s}}{\ave{\delta s^{2}}},\label{eq:fluct:CBS}
\end{align}
where we have expressed $C_{BS}$ also in terms of quark degrees of
freedom, noting that the baryon number of a quark is $1/3$ and the
strangeness of a s-quark is negative one. Here $(u,d,s)$ represent
the net-number of up, down and strange quarks, i.e., the difference
between up and anti-up quarks etc. For a non-interacting QGP, $\ave{\delta u\,\delta s}=\ave{\delta d\,\delta s}=0$,
so that $C_{BS}=1$. For a gas of kaons and anti-kaons, on the other
hand, where a light (up or down) quark is always correlated with a
strange anti-quark (kaons) or vice versa (anti-kaons) $\ave{\delta u\,\delta s}<0$,
resulting in $C_{BS}<1$. Strange baryons, on the other hand, correlate
light quarks with strange quarks or light anti-quarks with strange
anti-quarks, so that $\ave{\delta u\,\delta s}>0$. Therefore, for
sufficiently large values of the baryon number chemical potential,
$C_{BS}>1$ for a hadron gas, whereas for a non-interacting QGP $C_{BS}=1$
for all values of the chemical potential \cite{Koch:2005vg}. Since
$C_{BS}$ can be expressed in terms of susceptibilities, $C_{BS}=-3\frac{\chi^{B,S}_{11}}{\chi^{S}_{2}}$,
it can and has been calculated on the lattice with physical quark
masses by two groups \cite{Borsanyi:2011sw,Bazavov:2012jq}. Both
calculations agree with each other, and both report a small, but significant
difference between the lattice results and that from the hadron resonance
gas. In \cite{Bazavov:2014xya} it has been argued that this discrepancy
may be removed by allowing for additional strange hadrons, which are
not in the tables of the Particle Data Group (PDG) \cite{Agashe:2014kda},
but are predicted by various quark models. This is shown in Fig.~\ref{fig:fluct:cbs},
where the lattice QCD results are compared with a hadron resonance
gas based on all the hadrons in the Review of Particles \cite{Agashe:2014kda}
(dotted line) and a hadron gas with additional strange hadrons (full
line). Whether or not this turns out to be the correct explanation,
this comparison demonstrates that these cumulant ratios are a sensitive
probe of the relevant microscopic degrees of freedom.

To summarize, the above examples illustrate how cumulants of conserved
charges can be utilized to extract useful information about the correlations
and relevant degrees of freedom of QCD matter. Since they are amenable
to lattice QCD methods, the insights derived from such studies are
rather model independent.

\section{Measuring Cumulants}
\label{sec:3}

\subsection{Some general considerations}
Given the wealth of information which can be extracted from cumulants
of conserved charges and the fact that they can be determined model
independently, it would be desirable to measure these cumulants in
heavy ion collisions. However, a heavy ion collision is a highly dynamical
process whereas lattice QCD deals with a static system in global equilibrium.
In addition, real experiments have limitations in acceptance etc.,
which are difficult to map onto a lattice QCD calculation. Consequently
a direct comparison of experiment with lattice QCD results for fluctuation
observables is a non-trivial task. In the following we will discuss
various issues which need to be understood and addressed in order
for such a comparison to be meaningful.
\begin{itemize}
\item \textbf{Dynamical evolution:} So far our discussion assumed that the
system is static and in global thermal equilibrium. However, even
if fluid dynamics is applicable the system is at best in local thermal
equilibrium. The difference between local and global thermal equilibrium
is an important aspect of the evolution of fluctuations of conserved
charges, because the amount of conserved charge in a given co-moving
volume can only change by diffusion, and the rate of diffusion is
limited by causality \cite{Shuryak:2000pd}. This observation is central
to the use of the variable $R$ defined in Eq.~(\ref{R-def}) to
detect the presence of quark gluon plasma. If we consider a sufficiently
large rapidity window $\Delta y$ then the value of $R$ is frozen
during the QGP phase, and cannot change in the subsequent hadronic
stage. Of course, if $\Delta y$ is chosen too large, then $R$ never
equilibrates, and reflects properties of the initial state. This observation
can be made more quantitative using the theory of fluctuating hydrodynamics.
However, so far most theoretical studies have focused on schematic
models, see, for example \cite{Kitazawa:2013bta}.
\item \textbf{Global charge conservation:} Obviously, baryon number, electric
charge and strangeness are conserved globally, i.e., if we detected
all particles, none of the conserved charges would fluctuate. In contrast,
lattice QCD calculations are carried out in the grand canonical ensemble,
which allows for exchange of conserved charges with the heat bath.
Consequently, charges are conserved only on the average and, thus,
do fluctuate due to the exchange with the heat bath. These exchanges
and thus the fluctuations depend on the correlations between particles
and, as demonstrated above, on the magnitude of the charges of the
individual particles. Therefore, in order to compare with lattice
QCD, one has to mimic a grand canonical ensemble in experiment. This
can be done by analyzing only a subset of the particles in the final
state. However, even in this case, corrections due to global charge
conservation are present. These corrections increase with the order
of the cumulant \cite{Bzdak:2012an} and need to be taken into account
as discussed in \cite{Kitazawa:2013bta,Koch:2001zn,Aziz:2004qu,Kitazawa:2012at}. 
\item \textbf{Finite acceptance:} All real experiments do have a finite
acceptance, i.e., they are not able to cover all of phase space. In
addition, most experiments are unable to detect neutrons, which do
carry baryon number. However, due to rapid isospin exchange processes,
the lack of neutron detection may be modeled by a binomial
distribution \cite{Kitazawa:2012at}. While it is desirable to study
only a subset of particles, in order to mimic a grand canonical ensemble,
it is mandatory to have sufficient coverage in phase space in order
to capture all correlations.
\item \textbf{Efficiency corrections:} A real world experiment detects a
given particle only with a probability, commonly referred to as efficiency
$\epsilon$, which is smaller than one, $\epsilon<1$. However, this
does not imply that in every event the same fraction of produced particles
is detected. Rather, the number of measured particles fluctuates even
if the number of produced particles does not. In other words the finite
detection efficiency gives rise to fluctuations, which need to be
removed or unfolded before comparing with any theoretical calculation.
If the efficiency follows a binomial distribution, analytic formulas
for the relation between measured and true cumulants can be derived
\cite{Bzdak:2012ab,Bzdak:2013pha,Luo:2014rea}. Those have been applied
to the most recent analysis by the STAR collaboration. 
\item \textbf{Dynamical fluctuations:} A heavy ion collision is a highly
dynamical process and the initial conditions as well as the time evolution
may easily give rise to additional fluctuations. Especially at lower
energies, $\sqrt{s}\lesssim30\gev$, the incoming nuclei are stopped
effectively and deposit baryon number and electric charge in the 
mid-rapidity region. Clearly the amount of baryon number deposited will
vary from event to event, resulting in fluctuations of the baryon
number at mid-rapidity, which are not necessarily the same as those
of a thermal system. This potential source of background needs to
be understood and removed, especially at low energies where one uses
higher cumulants of the net proton distribution in order to find signals
for a possible QCD critical point. Not only does the number of baryon
and charges fluctuate due to the collision dynamics, so does the size
of the system. And while ratios of cumulants do not depend on the
average system size (unless the system is at a second order phase
transition), they are affected by event by event fluctuation
of the system size. This has been studied in \cite{Skokov:2012ds}
and it was found that only for the very most central collisions these
fluctuations are suppressed. Alternatively, one can device observables,
which are not sensitive to size fluctuation \cite{Jeon:2003gk,Koch:2008ia,Gazdzicki:2013ana,Sangaline:2015bma}. 
\end{itemize}
The first three points deserve some additional discussion, as they
pose contradictory demands on the measurement \cite{Koch:2008ia}.
In order to minimize corrections from global charge conservation one
wants to keep the acceptance window $\Delta$, say in rapidity, as
small as possible. On the other hand, in order to capture the physics,
the acceptance window needs to be sufficiently wide in order to catch
the correlation among the particles. Therefore, if $\sigma$ is the
correlation length in rapidity and $\Delta_{charge}$ the range over
which all the charges are distributed, then $\Delta/\Delta_{charge}\ll1$
in order to minimize the effects of charge conservation, and $\sigma/\Delta\ll1$
in order to capture the physics.

\begin{figure}[t]
\begin{center}
\includegraphics[width=0.6\textwidth]{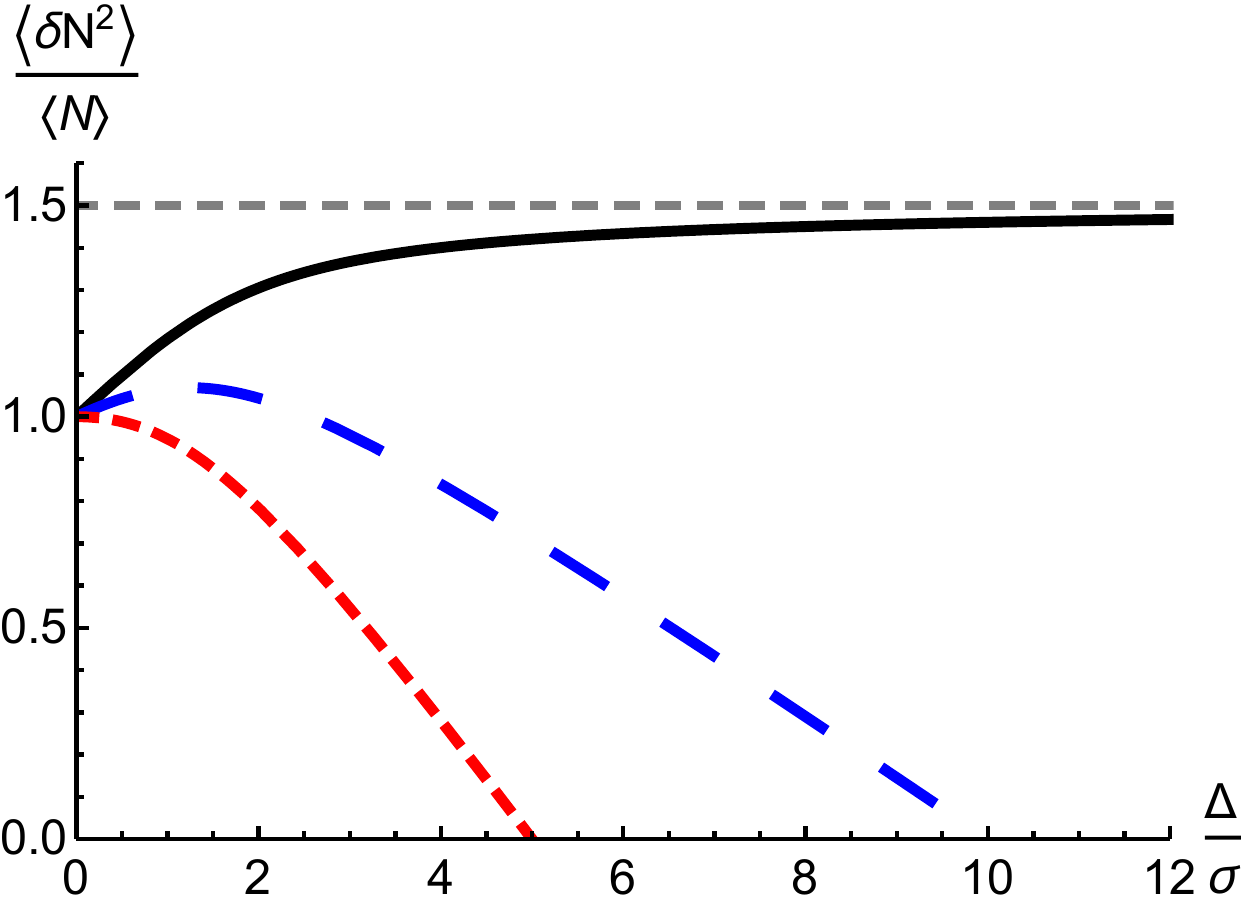} 
\end{center}
\caption{Observed scaled variance as a function of the acceptance window in
units of the correlation length. The full (black) line corresponds
to an infinite system where global charge conservation can be ignored.
The long-dashed (blue) and short-dashed (red) lines correspond to the
situation where the charge is conserved within $(-10\sigma,10\sigma)$
and $(-5\sigma,5\sigma)$, respectively.}
\label{fig:fluct:schematic_correlation} 
\end{figure}

To illustrate this point, let consider the following schematic model.
Let us define a two-particle correlation function (in rapidity $y$)
\begin{align}
\ave{n(y_{1})\left(n(y_{2})-\delta\left(y_{1}-y_{2}\right)\right)}=\ave{n\left(y_{1}\right)}\ave{n\left(y_{2}\right)}\left(1+C\left(y_{1},y_{2}\right)\right)
\end{align}
with $\ave{n(y)}=\rho=const.$ Then the (acceptance dependent) scaled
variance of the particle number is given by 
\begin{align}
\frac{\ave{(\delta N)^{2}}}{\ave N}=1+\frac{\rho}{\Delta}\int_{-\Delta/2}^{\Delta/2}dy_{1}\,dy_{2}\,C\left(y_{1},y_{2}\right)\label{fluct:eq:schematic_correlate}
\end{align}
where the acceptance in rapidity is given by $-\Delta/2<y<\Delta/2$.
Using a simple Gaussian for the correlation function 
\begin{align}
C\left(y_{1},y_{2}\right)=\frac{C_{0}}{\rho}\exp\left(-\frac{\left(y_{1}-y_{2}\right)^{2}}{2\sigma}\right)
\end{align}
in Fig.~\ref{fig:fluct:schematic_correlation} we show the scaled
variance as a function of the size of the acceptance window in units
of the correlation length $\Delta/\sigma$. The black line is simply
the expression of Eq.~(\ref{fluct:eq:schematic_correlate}), where
we have ignored any effects due to global charge conservation, i.e.,
$\Delta_{charge}\rightarrow\infty$. The red and blue dashed lines
represent the situation where the total charge is distributed over
a range of $\Delta_{charge}/\sigma\le5$ and $\Delta_{charge}/\sigma\le10$,
respectively. Here we used the leading order formulas of \cite{Bleicher:2000ek}
to account for charge conservation noting that a more sophisticated
treatment a la \cite{Sakaida:2014pya} would not change the picture
qualitatively. Lattice QCD and model calculations, on the other hand,
would give the asymptotic value indicated by the dashed gray line,
which we have chosen to be $\frac{\ave{(\delta N)^{2}}}{\ave N}=1.5$.
The obvious lesson from this exercise is that a comparison of a measurement
at one single acceptance window $\Delta$ with any model calculation
is rather meaningless. Instead, one needs to measure the cumulants
for various values of $\Delta$, and remove the effect of charge conservation.
If the subsequent results trend towards an asymptotic value for large
$\Delta$, it is this value which should be compared with model and
lattice calculations. Such a program has been carried by the ALICE
collaboration in order to extract the aforementioned charge fluctuations
\cite{Abelev:2012pv}. In this context it is worth mentioning that
recent comparisons of measured cumulant ratios with Lattice QCD to
extract the freeze out conditions \cite{Karsch:2012wm,Borsanyi:2013hza}
are based on measured cumulants which have not been extrapolated as
described above. Thus these results need to be taken with some care. 

\begin{figure}
\begin{center}
\includegraphics[width=0.45\textwidth]{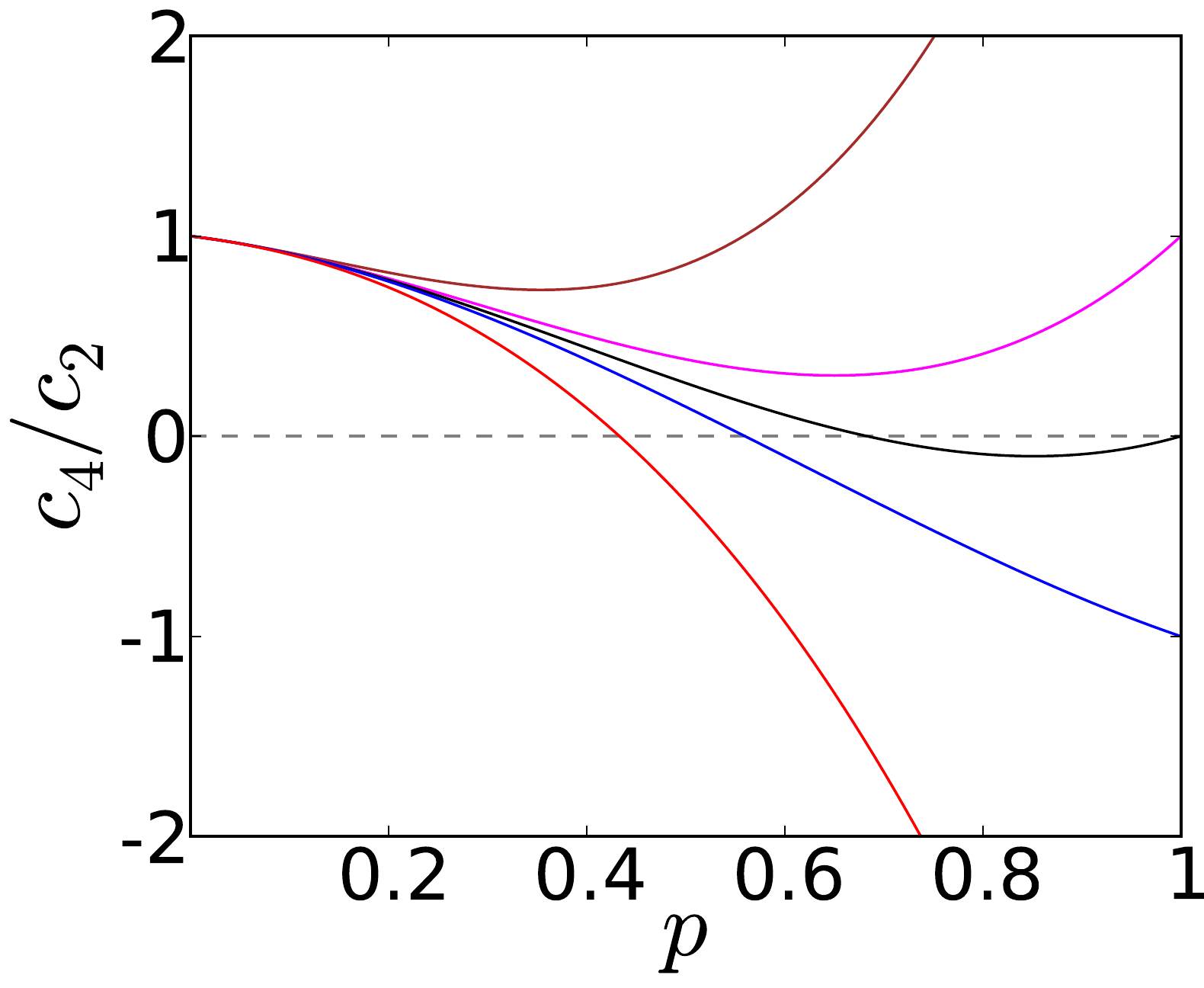} \hspace{1cm}
\includegraphics[width=0.45\textwidth]{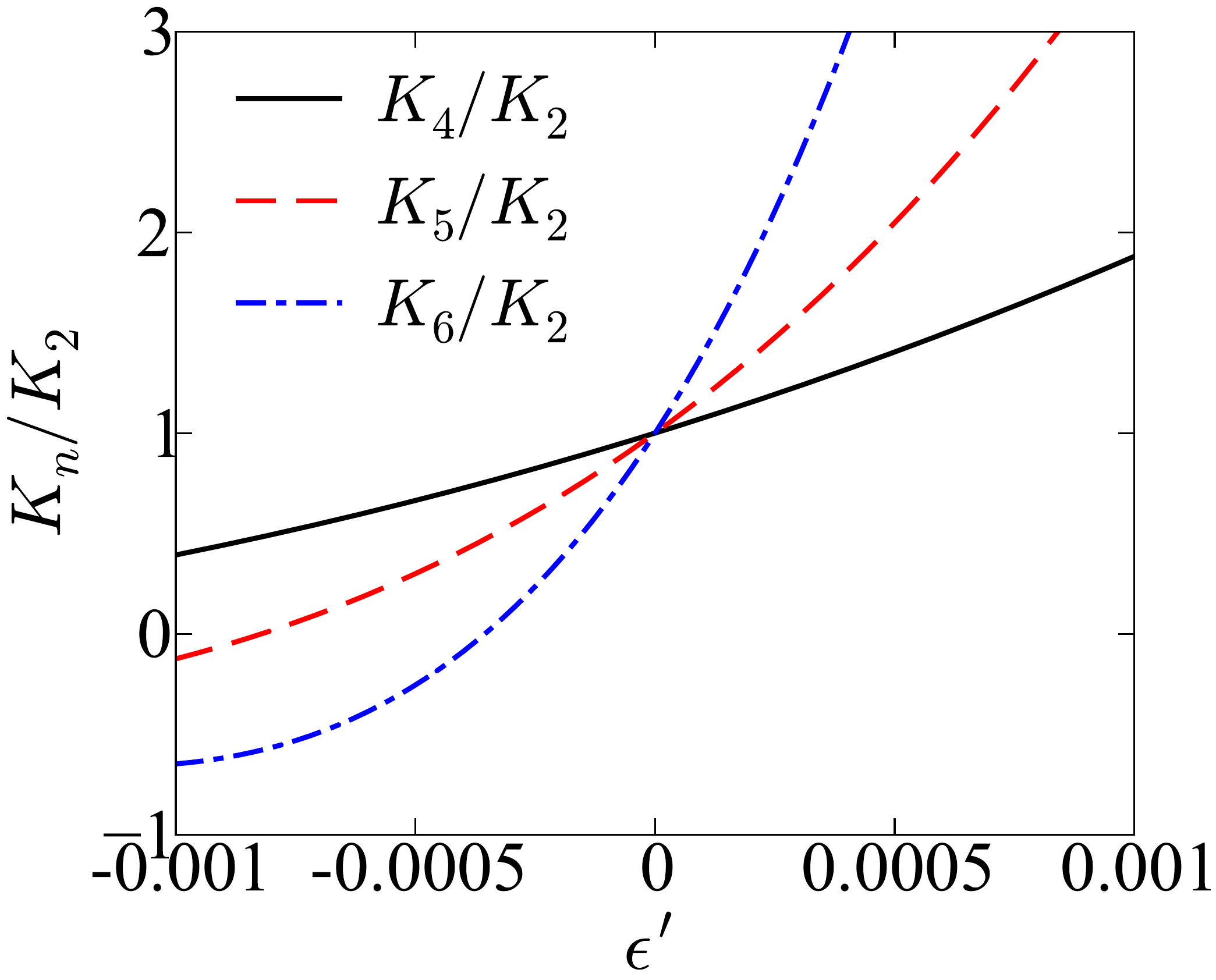}
\end{center}
\caption{Left panel: Observed cumulant ratio as a function of binomial probability
$p$. The lines from top to bottom correspond to true cumulant ratios
of $K_{4}/K_{2}=5,\,1,\,0,\,-1,\,-5$. Figure adapted from \cite{Bzdak:2012ab}.
Right panel: Effect of multiplicity dependent efficiency on various
cumulant ratios. Deviations form $K_{n}/K_{2}=1$ indicate the effect
of unfolding based on the binomial distribution with constant efficiency, 
$\epsilon_{0}$.  
For reference, the STAR data at $7.7\protect\gev$ exhibit a multiplicity dependence
corresponding to $\epsilon'\simeq -5\times 10^{-4}$ \cite{Luo:2015ewa}. Figure
adapted from \cite{Bzdak:2016qdc}.}
\label{fig:mult_eps}
\end{figure}

Before we turn to the proton cumulants, let us make a few additional
remarks concerning efficiency corrections, as they do play a significant
role in the recent STAR data \cite{Luo:2015ewa}. As can be seen in
the left panel of Fig. \ref{fig:mult_eps}, finite detection efficiency,
$\epsilon<1$, affects the observed cumulants considerably, and, thus,
needs to be corrected or unfolded. As discussed in Refs. \cite{Bzdak:2012ab,Bzdak:2013pha,Luo:2014rea}
such an unfolding can be done analytically if the probability for
detection of a particle follows a binomial distribution. However,
there is no a priori reason why the response of a complicated detector
should follow a binomial distribution. For example, in most experiments
the efficiency depends on the particle multiplicity, which would not
be the case for a binomial distribution where the binomial probability,
i.e, the efficiency, is constant, independent of the number of Bernoulli
trials. In Ref. \cite{Bzdak:2016qdc} the effect of a multiplicity
dependent efficiency and various other corrections have been explored.
In the right panel of Fig. \ref{fig:mult_eps} we show the resulting
cumulant ratios $K_{n}/K_{2}$ assuming that the efficiency depends
linearly on the multiplicity $M$
\[
\epsilon\left(M\right)=\epsilon_{0}+\epsilon'\left(M-\ave M\right).
\]
 Already a rather weak multiplicity dependence gives rise to correction
of order $50\%$ for $K_{4}/K_{2}$. 

The multiplicity dependence of the efficiency is just one example
for a non-binomial behavior of the detection probability. There are
certainly others and some examples are discussed in \cite{Bzdak:2016qdc}.
Therefore, the only way to assure that detector effects are probably
accounted for is for individual experiments to simulate and understand
the response of the detector and carry out the necessary unfolding.
That such an exercise is necessary should be obvious from the above
examples.

\subsection{Proton cumulants}

Let us next turn to the net-proton cumulants. It has been suggested
that higher order baryon-number cumulants are particularly sensitive
to the presence of a critical point in the QCD phase-diagram \cite{Stephanov:2008qz}.
Since it is difficult to detect neutrons, this let to a series of
measurements of net-proton cumulants at various energies \cite{Luo:2015ewa,Adamczyk:2013dal}.
As shown in \cite{Kitazawa:2012at}, given rapid, pion-driven, isospin
exchange, the absence of neutrons can be rather reliably modeled by
a binomial process, with binomial probability $p\simeq0.5$. This,
on the other hand, implies that in addition to the detection efficiencies
one also needs to unfold the absence of neutrons, or, in other words,
detection of protons with an efficiency of 0.8 corresponds to detection
of baryons with an effective efficiency of 0.4. As a result, the sensitivity
to the correct magnitude of the true cumulants gets considerably reduced
as can be seen in the left panel of Fig. \ref{fig:mult_eps}. 

\begin{figure}
\begin{center}
\includegraphics[width=1.0\textwidth]{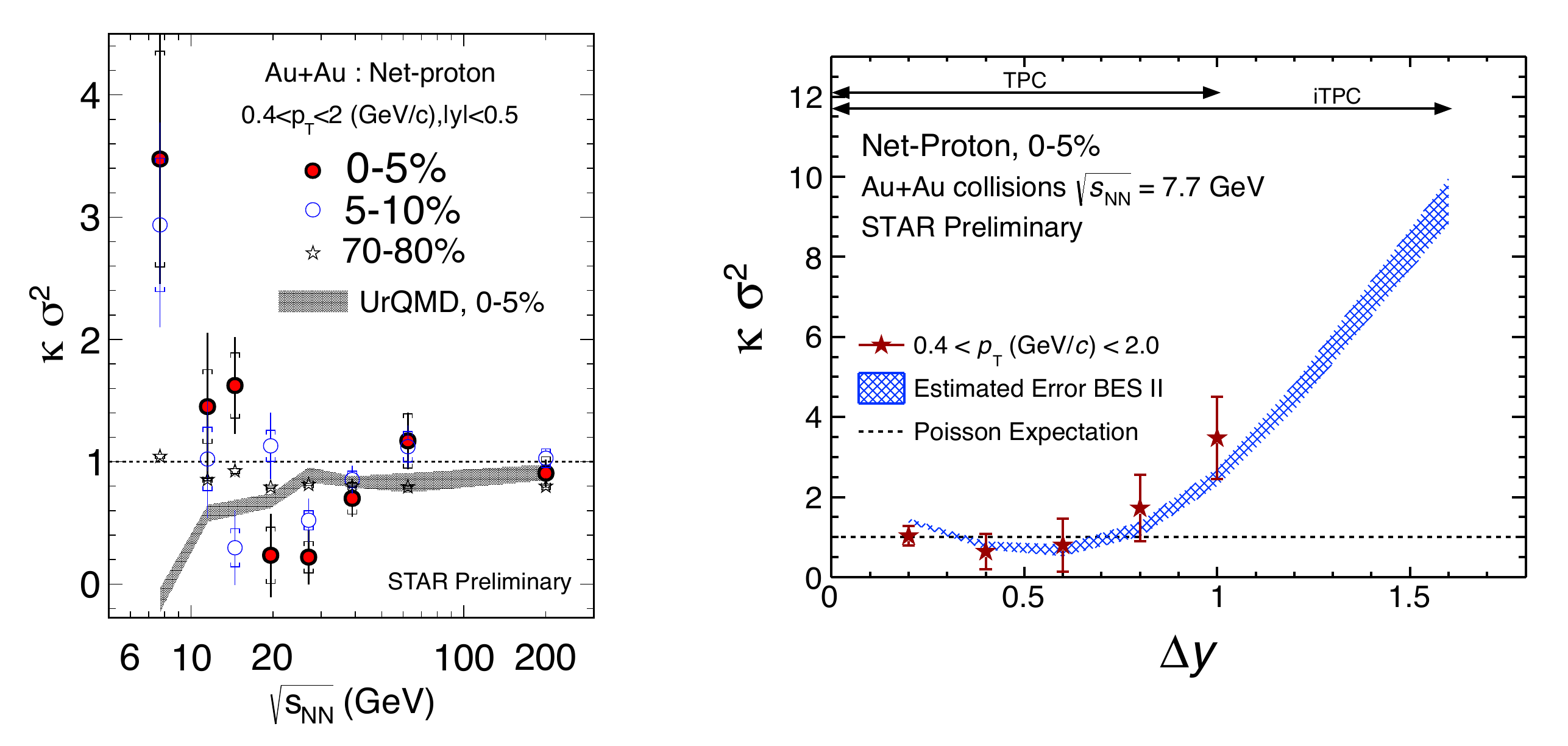} 
\end{center}
\caption{Preliminary data by the STAR collaboration for the energy and rapidity
dependence of the $K_{4}/K_{2}$ cumulant ratio. Figures adapted from
\cite{Luo:2015ewa,Luo:2015doi}.}
\label{fig:star_k4}
\end{figure}

Finally let us discuss the preliminary results for the $K_{4}/K_{2}$
cumulant ratio for net-protons obtained by the STAR collaboration.
In Fig. \ref{fig:star_k4} we show both the dependence on the beam
energy (left panel) and on the width of the rapidity window (right
panel) for the lowest beam energy of $\sqrt{s_{NN}}=7.7\gev$. Also
shown in the left panel are results from UrQMD calculations. These
exhibit a decreasing cumulant ratio with decreasing beam energy, which
is due to baryon number conservation \cite{Bzdak:2012an}. This behavior
is in stark contrast with the measured cumulant ratio, which shows
a steep rise towards lower energies. It is noteworthy, that this rise
only occurs $after$ corrections for efficiency based on a binomial
efficiency distribution have been applied \cite{Luo:2015ewa}. Obviously,
these preliminary data are very intriguing, especially since most
``trivial'' effects tend to reduce the cumulant ratios, such as
the aforementioned baryon number conservation. However, in light of
our discussion, it will be important that the validity of the binomial
efficiency distribution is verified by a detailed analysis of the
STAR detector response.

The dependence on the size of the rapidity window, shown in the right
panel of Fig. \ref{fig:star_k4} is also quite interesting. The cumulant
ratio keeps increasing up to the maximum available range of $\Delta y=1$.
Following our simple model consideration, this implies the the underlying
rapidity correlations are rather long range. Typically long range
rapidity correlations are associated with early times in the collision.
Although this correspondence is somewhat washed out at lower beam
energies, it raises the question if the observed signal may be due
to some initial state effects such as impact parameter/volume or stopping
fluctuations.

\section{Discussion}

In this contribution we have discussed fluctuations of conserved charges
and their utility for the exploration of QCD matter. In particular
we have concentrated on various cumulants which have the advantage
of being accessible to Lattice QCD calculations. Alternatively one
may study the underlying correlations, as suggested by Bialas et al.
\cite{Bialas:1999tv}. These may actually provide more physical insight
into the dynamics at play. If only one particle species is being considered,
such as e.g. protons, one can relate the cumulants and the correlation
functions \cite{Ling:2015yau,bzdak2016}. For example the two-particle
correlation function is simply given by the first and second order
cumulants, $K_{1,}\,K_{2}$,
\[
C_{2}=K_{2}-K_{1}.
\]
However, once net-protons, i.e. protons \emph{and} anti-protons, are
being considered, no direct relations between the correlation functions
and the cumulants exist. Instead additional (factorial) moments are
required, which can be measured but not be calculated in Lattice QCD. 

To conclude, fluctuations are a powerful tool to explore the structure
of QCD matter. They provide insight into the relevant degrees of freedom,
their correlations, and possible phase structures. The measurement
of fluctuations requires some care. First the detector response needs
to be well understood and removed by a proper unfolding procedure.
Furthermore, since a heavy ion collision is a highly dynamical process,
fluctuations induced by the initial state or by the dynamical evolution
need to be understood before a comparison with model or Lattice QCD
calculations is possible. Preliminary data on net-proton cumulants
from the STAR collaboration show intriguing features, especially at
the lowest energies. To which extend they constitute our first glimpse
at structures in the QCD phase diagram can only be found out if all
these uncertainties are fully understood.

\section{Acknowledgments}
This work is supported by the Director, Office of Energy Research,
Office of High Energy and Nuclear Physics, Divisions of Nuclear Physics,
of the U.S. Department of Energy under Contract No. DE-AC02-05CH11231, and 
by the Polish Ministry of Science and Higher Education (MNiSW), 
by founding from the Foundation for Polish Science, 
and by the National Science Centre (Narodowe Centrum Nauki), 
Grant No. DEC-2014/15/B/ST2/00175 and in part by DEC-2013/09/B/ST2/00497.

\end{document}